\documentclass[lettersize,journal]{IEEEtran}
\pdfoutput=1
\usepackage{amsmath,amsfonts}
\usepackage{algorithmic}
\usepackage{algorithm}
\usepackage{array}
\usepackage[caption=false,font=normalsize,labelfont=sf,textfont=sf]{subfig}
\usepackage{textcomp}
\usepackage{stfloats}
\usepackage{url}
\usepackage{verbatim}
\usepackage{graphicx}
\usepackage{cite}
\usepackage{amsmath}
\usepackage{amssymb}
\usepackage{mathtools}
\usepackage{amsthm}

\usepackage{wrapfig,lipsum,booktabs}
\usepackage{hyperref}
\usepackage{url}
\usepackage{sidecap}
\usepackage{amsmath,bm}
\usepackage{multirow}
\usepackage{graphicx}
\usepackage{comment}
\usepackage{amssymb}
\usepackage{commath}
\usepackage{mathtools, nccmath}
\usepackage{enumitem}
\usepackage{color, colortbl}

\usepackage[capitalize,noabbrev]{cleveref}

\theoremstyle{plain}

\theoremstyle{definition}

\theoremstyle{remark}

\newcommand{\hq}[1]{\textcolor{black}{#1}}
\newcommand{\gl}[1]{\textcolor{black}{#1}}

\newcommand{\yhq}[1]{\textcolor{black}{#1}}
\newcommand{\lin}[1]{\textcolor{black}{#1}}

\newcommand{\hanqi}[1]{\textcolor{black}{#1}}

\newcommand{\rev}[1]{\textcolor{black}{#1}}

\begin{document}

\title{Explainable Recommender with Geometric Information Bottleneck}

\author{Hanqi Yan, Lin Gui, Menghan Wang, Kun Zhang, Yulan He
\thanks{Manuscript received March 23, 2023; revised November 07, 2023, accepted December 23, 2023.}}

\maketitle

\begin{abstract}
Explainable recommender systems can explain their recommendation decisions, enhancing user trust in the systems.
Most explainable recommender systems either rely on human-annotated rationales to train models for explanation generation or leverage the attention mechanism to extract important text spans from reviews as explanations. The extracted rationales are often confined to an individual review and may fail to identify the implicit features beyond the review text. To avoid the expensive human annotation process and to generate explanations beyond individual reviews, we propose to incorporate a geometric prior learnt from user-item interactions into a variational network which infers latent factors from user-item reviews.
The latent factors from an individual user-item pair can be used for both recommendation and explanation generation, which naturally inherit the global characteristics encoded in the prior knowledge.
Experimental results on three e-commerce datasets show that our model significantly improves the interpretability of a variational recommender using the Wasserstein distance while achieving performance comparable to existing content-based recommender systems in terms of recommendation behaviours.  The code is publicly available at \url{https://github.com/hanqi-qi/Explainable-Recommender-GIANT}.
\end{abstract}

\begin{figure*}[t]
    \centering
    \includegraphics[width=0.92\textwidth,trim={6 2 10 9},clip]{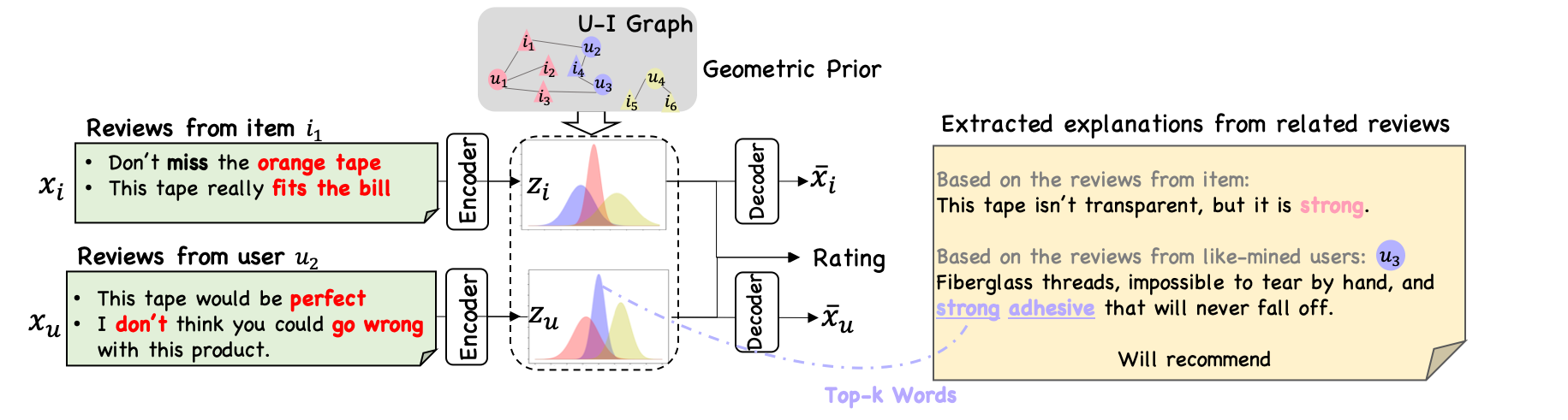}
    \caption{The proposed geometric prior reflects the U-I distances in their interaction graph, frequently interacted users/items are assigned into the same cluster (illustrated in node colour). Existing extractive methods can only identify indicative words (shown in red) from the given review pair $(x_i,x_u)$, \hanqi{while our method can extract sentence-level reviews written by other like-minded users or on similar items, which are assigned into the same latent dimension/cluster, i.e., we extract the sentences from the review by user $u_3$ which is in the same cluster as user $u_2$}, as the reviews of $u_2$ are too general to summarise the key features.}
    \label{fig:intro}
\end{figure*}

\section{Introduction}
Typically, a recommender system compares users' preferences with item characteristics (e.g., item descriptions or item-associated reviews) or studies user-item historical interactions (e.g., ratings, purchases or clicking behaviours) in order to identify items that are likely of interest to users. In addition to item recommendation, interpretable recommenders also generate the rationale behind the recommendation decision 
\cite{ghazimatin2020prince,zhang2020explainable}. Most existing interpretable recommenders can either generate rationale or extract text spans from a given user-item review as explanations of model decisions. Rationale generation requires annotated data for training, 
e.g., short comments provided by users explaining their behaviours of interacting with items. \hanqi{The extractive explainable recommenders either rely on}
annotated sentiment-bearings aspect spans in reviews for training \cite{DBLP:conf/sigir/ZhangL0ZLM14,DBLP:conf/emnlp/NiLM19,DBLP:conf/ijcai/ChenW0WBWC19,10.1145/3340531.3411992,10.1145/3459637.3482420} \hanqi{or extract sentences from a target item's review as recommendation rationales based on attention weights 
\cite{DBLP:conf/www/ChenZLM18,DBLP:conf/www/WangCW22,DBLP:conf/naacl/WuWLH19,DBLP:conf/kdd/TayLH18,DBLP:conf/sigir/LiQPQDW19,DBLP:conf/aaai/ChenZQ19,DBLP:journals/tois/LiuWPWWJ21,zhang2022aenar}.}

We argue that explanation generation models are often supervised by human-annotated rationales, which are labour-intensive to obtain in practice. \hanqi{The explainable recommenders built on extractive approaches can only extract rationales from an individual user-item review document. They suffer from the following limitations.}
First, some reviews may be too general to explain the user rating, rendering them useless for explanation generation. 
For example, the review `\emph{I really like the smartphone, will recommend it to my friends}' does not provide any clue why the user likes the smartphone. Second, features directly extracted from a review document may fail to reflect some global properties which can only be identified from implicit user-item interactions. Meaningful insights could still be derived from reviews towards items that are not directly purchased/rated by a user but preferred by other like-minded users. 

To address the aforementioned limitations, we propose a \textbf{G}eometric \textbf{I}nform\textbf{A}tio\textbf{N} bo\textbf{T}tleneck~(\texttt{\textbf{GIANT}}) framework to incorporate the prior learned from a user-item interaction graph to refine the induced latent factors learnt from reviews. Essentially, we use the prior learned from one modality (i.e., the user-item interaction graph) to constrain the learning of latent factors in another modality (i.e., the review text). In this way, we naturally address the problem of the over-simplified Gaussian prior assumption in approaches built on the Variational Autoencoder (VAE)~\cite{DBLP:journals/corr/KingmaW13}. More concretely, our prior is derived from the user/item clusters obtained from the user-item interaction graph. Each user or item has its soft cluster assignment represented as a probability distribution over clusters, which is in proportional to its distance to each cluster centroid. For this reason, we call our prior the geometric prior as it translates the geometric properties of users/items into their cluster-associated probability distributions. The prior is subsequently incorporated into the variational network for constraining the learning of latent factors from the review text. For a user-item pair, $(u, i)$, we feed the user's past reviews and the item's associated reviews (excluding $u$'s review on $i$) to the variational network for the inference of user- and item-associated latent factors, which are subsequently used for rating prediction and explanation generation. 

Figure~\ref{fig:intro} illustrates how the proposed geometric prior can be used to develop the explainable recommender. 
To impose the prior on the latent variables of the variational framework taking the set of user reviews and item reviews as input, we apply the KL divergence to minimise the discrepancy between the posterior distribution of the user latent factor \hq{$z_u$}, and the item latent factor \hq{$z_i$}, learned from review text, and the geometric prior derived from the user-item interaction graph. 
The latent factors inferred by the variational network can be treated as topics summarising the key semantics of input reviews. For an input user-item pair, the most prominent  dimension of the user (or item) latent factor is considered as the most important topic revealing the user preference (or item characteristic). As will be shown in the experiments section, for each topic (i.e., a dimension of the latent factor), we can retrieve its top-$k$ associated words as its representation. To generate sentence-level explanations, we identify sentences from reviews residing in the same user/item cluster which are most similar to the user- and item-topic representations.~\footnote{Details about how to extract sentence-level review as explanations can be found in \S\ref{sec:gen_ex}: Explanation Generation.}  

Experimental results on the three commonly used benchmarking datasets show that the proposed method achieves performance comparable with several strong baselines in recommendation prediction. Moreover, both quantitative and qualitative analysis shows \hanqi{we can see clear and meaningful clusters distributed in the latent space regularised by the geometric prior,} \hanqi{when compared with VAEs using the standard Gaussian prior or employing the Wasserstein distance. The user study demonstrates that our generated explanations are better in \emph{relevance}, \emph{faithfulness} and \emph{informativeness}.}
\hanqi{Our contributions can be summarized below:}
\begin{itemize}[noitemsep,topsep=0pt]
    \item We proposed a geometric prior derived from user-item interactions in order to incorporate the global features from similar users/items, allowing recommendation explanations to be generated beyond explicit user-item interactions. 
    \item We give a proof that the geometric prior can be effectively incorporated into the variational framework with an upper bound in mutual information between the textual input and the latent representations regularized by the geometric prior.
    \item Experimental results show the efficiency of our proposed method in both review rating prediction and explanation generation.
\end{itemize}
\vspace{-2mm}
\section{Related Work}
We review recommender systems, with particular attention to 
those built on VAE or offering explanations of recommendation decisions. 
\paragraph{Recommender with VAE}
Text data such as user reviews, item or brand descriptions could be important for developing a high-quality recommender system as they can be exploited to address the sparsity issue in user-item interactions. HFT~\cite{10.1145/2507157.2507163} and CTR~\cite{DBLP:conf/kdd/WangB11} adopted Latent Dirichlet Allocation (LDA)~\cite{DBLP:journals/jmlr/BleiNJ03} to extract latent topics from review text. 
VAE can also be used for text modeling due to its ability in extracting latent and interpretable features~\cite{DBLP:conf/www/0001RWLJ21,DBLP:conf/wsdm/TruongSL21,DBLP:conf/sigir/WangJZ0XC20}. \cite{DBLP:conf/wsdm/TruongSL21} argued that the commonly used isotropic Gaussian in VAE is over-simplified and proposed BiVAE by introducing constrained adaptive prior for learning user- and item-dependent prior distributions. 
More recently, review information is used to disentangle the
latent user intents at the finer granularity, for example, DisenGCN~\cite{DBLP:conf/icml/Ma0KW019} and DisenHan~\cite{DBLP:conf/sigir/WangJZ0XC20} used the graph attention mechanism to differentiate multiple relations and features. 

\paragraph{Explainable Recommender System}
One popular interpretable method is to train a language generation model with the ground-truth explanations supplied, which can be the first sentence of a given review or human-annotated text spans in 
the review text~\cite{
DBLP:conf/emnlp/NiLM19,DBLP:journals/corr/abs-2202-07371}.
Some of these approaches firstly extracted user-aspect-opinion and item-aspect-opinion tuples as ground truth, then modelled the interactions between the two tuples to derive the rating and the indicative features~\cite{
DBLP:conf/cikm/TanXG00Z21,DBLP:conf/sigir/ChenYYH0020}. 
 \yhq{Retrieval-based (aka. extractive) solutions directly select representative text spans (rationales), i.e., words or sentences, from a target item's past reviews to explain the recommender behaviours. 
 Attention-based methods are prevalent in review-based explainable recommenders~\cite{
 DBLP:journals/tois/LiuWPWWJ21,zhang2022aenar}. Beyond leveraging attention weights to select rationales, ~\cite{
 DBLP:conf/www/YangWCDW22,pan2022accurate} proposed different optimization objectives to refine the selection. }
 
However, the aforementioned approaches suffer from two major issues.
\yhq{Some approaches require aspects to be extracted first in order to provide aspect-related predictive features as interpretations, thus heavily dependent on the quality of the aspect extractors employed; while others rely on human-annotated rationales to train supervised language generators for explanation generation. Our proposed approach can instead generate explanations with only supervision signals coming from review ratings.}

\vspace{-2mm}
\section{Geometric Information Bottleneck}
\label{sec:GIB}
To incorporate the \hanqi{global features from actively interacted users/items to explain the recommendation behaviour}, we propose to incorporate the geometric prior derived from user-item interaction graphs to learn the latent factors of users and items from review text in a variational network. Although various work has been proposed to adopt richer priors to infer a more complex and realistic posterior in standard VAEs~\cite{DBLP:journals/corr/TomczakW17,DBLP:journals/corr/ZhaoKZRL17}, our solution is different from existing approaches as we need to impose the prior knowledge learned from the graph modality to constrain the variables in the text modality. In what follows, we show how this can be done under the theory of information bottleneck. 

\yhq{Based on the Information Bottleneck (\textsl{IB}) theory ~\cite{https://doi.org/10.48550/arxiv.1503.02406}, we can train an encoder-decoder architecture $x \xrightarrow[]{encode} z \xrightarrow[]{decode} \Bar{x}$, where $x \in X$, $z \in Z$, $\Bar{x} \in \Bar{X}$ are the input, the hidden, and the reconstructed representations, respectively, in order to preserve the meaningful information about $\bar{X}$ in $Z$ while maximally compress the information of $X$. We achieve this by maximising the following:
}
\begin{equation}\small
    \lin{O_{IB} = I(\Bar{X};Z) - \beta \cdot I(X;Z), }
    \label{eq:IB}
\end{equation}
where $I(\cdot)$ denotes the mutual information,
$\beta$ is a Lagrange multiplier. 
The second term is to maximally compress $X$, equivalent to minimising the mutual information between $X$ and $Z$, denoted as $I(X;Z) \leq I_c$, $I_c$ is the upper bound. 


Intuitively, we assume the input has representations $x^t \in X^t$ and $x^g \in X^g$ under different modalities\footnote{In the rest of the paper, we use superscript $t$ for text and $g$ for the graph.} satisfying the constraint, $I(X^t;X^g) \geq I_x$, since $X^t$ and $X^g$ are derived from the same input, they should be relevant. 
Then, two independent well-trained encoder-decoder architectures in the two modalities optimised by Eq. (\ref{eq:IB}) are constrained by $I(X^t;Z^t) \leq I_c$ and $I(X^g;Z^g) \leq I_c$, respectively. \yhq{We derive a new upper bound for the multi-modality IB as follows:}
\footnote{The proof is shown in the Appendix B.}
\begin{equation}\small
    I(X^t;Z^g) \leq H(X^t) - H(X^g) + H(X^g|X^t) + I_c ,
    \label{eq:transIB}
\end{equation}
where $H(\cdot)$ and $H(\cdot|\cdot)$ are entropy and conditional entropy, respectively. 
Therefore, we can maximise the objective function using the input from the text modality, with a regularisation from another modality, 
\footnote{The proof is shown in the Appendix B.}
\begin{equation}\small
    \lin{O^t_{IB} = I(\Bar{X}^t;Z^t) - \beta \cdot I(X^t;Z^g).}
    \label{eq:O_ib}
\end{equation}
According to \cite{DBLP:conf/iclr/AlemiFD017}, the above equation can be approximated by applying the reparameterisation trick \cite{https://doi.org/10.48550/arxiv.1312.6114} with a random Gaussian noise $\epsilon$:
\begin{equation}\small
\begin{split}
  O^t_{IB} &=  \mathbb{E}_{x^t\sim X^t}  \mathbb{E}_{\epsilon \sim p(\epsilon)}{[ - \underbrace{{\rm log}q(\Bar{x}^t|f(x^t,\epsilon)]}_{{\rm Recons. \; term:\;}{Q(x^t)}} } \\
  & +\beta \cdot\underbrace{ {\rm KL}(p(Z^t|x^t,\epsilon)|r(Z^g))}_{{\rm KL-div. \; term: \;} {\rm KL}(Z^t|Z^g)},
  \label{eq:obj_Lagrange}
\end{split}    
\end{equation}
where $x^t$ is one training sample from input $X^t$, $\Bar{x}^t$ is the reconstruction output, $q(\cdot)$ and $p(\cdot)$ are the posterior probabilities estimated by the decoder and the encoder, respectively, $r(Z^g)$ is the estimated distribution from user-item interaction graph. Here, Eq. (\ref{eq:obj_Lagrange}) has a similar form as $\beta$-VAE which contains a reconstruction term $Q(x^t)$ defined by the posterior probability and a KL-divergence term ${\rm{KL}}(Z^t|Z^g)$ regularised by the prior from another modality $Z^g$. 

\hanqi{However, two key challenges need to be addressed when incorporating the geometric prior into a variational network: (1) How to derive prior $r(Z^g)$ from the user-item interaction graph; (2) How to compute $Q(x^t)$, which is defined as the log conditional probability of ${\rm log}q(\Bar{x}^t|f(x^t,\epsilon))$ in Eq. (\ref{eq:obj_Lagrange}).}

\paragraph{Geometric prior obtained from clustering with Gaussian kernel.}
\label{sec:graph_prior}
\hanqi{
We now discuss how to derive the geometric prior from the user-item interaction graph, 
in which a node can be either a user or an item and the frequently interacted nodes are more likely to share similar preferences or characteristics.} We can use a graph neural network to learn user/item node representations \hanqi{with review ratings as the supervision}. Afterwards, \hanqi{similar users/items are naturally gathered into one of the $K$ clusters,} denoted as $\bm{C}_{k}$, $k\in \{1,2,\cdots, K\}.$~\footnote{\hanqi{In the following, We omit the sbscript $u$, $i$ for the user and item as they are processed in the same way.}} 

We next show how to define the geometric prior based on the clustering results. 
Here we use the Gaussian kernel to compute the distance $\rho^{g}_{k}$ between a user or item and the centroid of its corresponding cluster $\bm{C}_k$ as in Eq. (\ref{eq:cluster}). 
\begin{equation}\small
    \rho^g_{k} = \frac{\exp\big(-\norm{\bm{e}^g-\bm{C}_{k}}^{2}/2\alpha^{2}\big)}{\sum_{k^{'}=1}^{K}\exp\big(-\norm{\bm{e}^g-\bm{C}_{k'}}^{2}/2\alpha^2\big)},
    \label{eq:cluster}
\end{equation}
\noindent where $\bm{e}^g$ is graph node embedding for user or item, $\bm{C}_{k}$ is the corresponding $k$-th cluster centroid vector, $\alpha$ is a hyper-parameter used to adjust the distribution according to the data density. 

It can be easily shown that $\Sigma_{k} \rho^g_{k} = 1$. That is, the distribution of $\{\sqrt{\rho^g_{k}}\}$ resides in a hyperspherical cap area on a uniform $K$-ball, weakly approximating to a Gaussian distribution when $K$ is large~\cite{AIHPB_1987__23_S2_397_0}. We use $r(Z^g)$ to denote the user's or item's probability distribution over clusters obtained from the user-item interaction graph, i.e., $r(Z^g)=\{r(z_u^g), r(z_i^g)\}$, where $r(z_u^g)=[\rho^g_{u1},\cdots,\rho^g_{uK}]$ and $r(z_i^g)=[\rho^g_{i1},\cdots,\rho^g_{iK}]$. 

\paragraph{Posterior over conditional probability.}
The reconstruction term $Q(x^t)$ in Eq. (\ref{eq:obj_Lagrange}) can be derived by the Nadaraya-Watson estimator \cite{10.2307/2669691}:
\begin{align}\small
    Q(x^t) &= {\rm log}\frac{\hat{p}(\Bar{x}^t_{\epsilon},x^t)}{\hat{p}(x^t)} \nonumber\\
    &= {\rm log}\frac{\frac{1}{N} \sum_{j=1}^N \kappa(\frac{x^t-x^t_j}{h}) \cdot \kappa(\frac{\Bar{x}^t_{\epsilon}-x^t_j}{h})}{\frac{1}{N} \sum_{j=1}^N 
    \kappa(\frac{x^t-x^t_j}{h}) },
\end{align}
where $\kappa(\cdot)$ is a kernel function for density estimation, $\Bar{x}^t_{\epsilon}$ is the reconstructed $x^t$ with random noise $\epsilon$, and $x^t_j$ is the $j$-th training sample. In practice, we choose the RBF kernel $\kappa = {\exp}(-||x-x'||^2))$ and ignore the denominator since the goal is to optimise the decoding of $\Bar{x}^t$. Then we are able to obtain the following bound by the triangle inequality in the learned metric space as $Q(x^t)  \varpropto {\rm log(exp}(-||\Bar{x}^t_{\epsilon} - x^t||^2))$.
By minimising the loss, we can preserve the key local similarity of nearby representations while ensuring different representations from different feature characteristics~\cite{DBLP:conf/nips/CzolbeKCI20,DBLP:journals/tip/WangBSS04}. 

Therefore, we are able to guarantee that the information bottleneck with our proposed geometric prior is:
\begin{itemize}[noitemsep,topsep=0pt]
    \item \textbf{Transferable} $-$ According to Eq. (\ref{eq:transIB}), the \hanqi{mutual information between textual input and the compressed latent variable regularized by geometric prior} is still upper bounded by the global information bottleneck, i.e., we can take the bound from one modality to guide the training on the other.
    \item \textbf{Practicable} $-$ 
    With our solution in Eq. (\ref{eq:cluster}), we do not have to utilise a pre-defined prior or oversimplified prior, but instead use the \hanqi{Gaussian-kernel} based distance distribution to regularize the latent space. \hanqi{The empirical recommendation results in Table~\ref{tab:main_results} shows the superiority of the geometric prior.}
    \item \textbf{Interpretable} $-$ The alignment between representations from different modalities and their geometric space based probabilities makes it possible to use latent factors inferred from review text, \hanqi{i.e., the semantic clusters embedded in various latent dimensions as shown in Table~\ref{tab:our_clusterwords}, to provide recommendation explanation.}
\end{itemize}
\vspace{-1.5mm}
\section{The GIANT Framework}
\hanqi{In this section, we provide a practical implementation of our geometric information bottleneck and use it to achieve item recommendation and explanation generation.}

\paragraph{Clustering on Graph Node Embedding.}
We construct a user-item interaction graph where nodes are users or items, and edges are created if the corresponding user's rating on the item is above the item's average rating. We then use \textsl{LightGCN}~\cite{DBLP:conf/sigir/0001DWLZ020} 
to obtain the user/item embeddings. \rev{This choice is motivated by its omission of feature transformation and nonlinear activation, making it a popular choice in recent successful recommenders, as demonstrated by studies such as ~\cite{mao2021ultragcn,peng2022svd,WANG2023119465}}.
Afterwards, we apply K-Means on the learned graph node embeddings to derive the cluster centroid vectors.

\paragraph{Textual Feature Extractor.} For a given user-item pair, $(u,i)$, the input to our recommender is a set of reviews written by $u$ and a set of reviews on $i$ (excluding $u$'s review on $i$). As each user or item is associated with multiple reviews, and some reviews might be very long, the Convolutional Neural Network (CNN) is used to encode reviews due to its efficiency in encoding long sequences, as has been studied in \cite{DBLP:conf/www/ChenZLM18}. 
We then aggregate multiple users/items reviews by the attention mechanism to derive the final textual vectors,  denoted as $x_u^t$ or $x_i^t$, respectively. Given the textual vectors, we employ two geometric regularised variational networks (i.e., the encoder-decoder structure) with shared weights to infer the latent factors, $z^t_u$ and $z^t_i$, obtained from the user and item review text, respectively.\footnote{In what follows, we drop the subscript $u$ and $i$ for simplicity.} 

\paragraph{Minimising Distribution Discrepancy.} 
To minimise the KL-divergence in Eq. (\ref{eq:obj_Lagrange}), we need to project $Z^g$ and $Z^t$ to a metric space with the same size. 
Specially, \gl{we obtain the latent variables $z^t = f(x^t,\epsilon)$ by applying the reparameterisation trick \cite{https://doi.org/10.48550/arxiv.1312.6114}.
However, due to the randomness in initialisation, the learned factors $Z^t$ might not find its correspondence in $Z^g$. Therefore, in our one-layer MLP based encoder, we propose a prior-centralisation trick which builds the connection between the clusters, obtained from the user-item interaction graph, and the encoder by making the encoder weights close to the cluster centroids $\bm{C} \in \mathbb{R}^{d\times k}$ obtained from the graph:}\useshortskip
\begin{equation}
        \mathcal{R}_{centroid} = \sum_{k=1}^K \norm{W_{en}^k - \bm{C}_k}^2,
\end{equation}
where $W_{en}^k$ denotes the $k$-th column of the encoder weight matrix $W_{en}$ 
and $\bm{C}_k$ denotes the representation of the $k$-th cluster centroid vector.
\gl{Note that $R_{centroid}$ will force the distribution of $Z^t$ approximate to the prior and ignore the input. Therefore, in practice, this regularisation is only deployed in the initial training epochs.}

Once the connection between the encoder weights and cluster centroids is established, the learning of the text latent variable $z^t$ can be regarded as a soft assignment of the text representation $x^t$ to various clusters.
Inspired by~\cite{van2008visualizing}, a Student $t$-distribution is used to mitigate the crowding problem. Here, we use $\tau$ to adjust the distribution tail and map its values to the range of $[0,1]$. The $k$-th dimension of a text latent variable $z^t_{k}$ denotes the probability of the corresponding user or item being assigned to the $k$-th cluster:
\begin{equation}\small
    \eta^t_{k} = \frac{1+e^{\varphi(z^t_{k})/\tau}}{\sum_{k'=1}^{K}e^{\varphi(z^t_{k'})/\tau}}, 
\label{eq:qz_softmax}
\end{equation}
\noindent \rev{where $\tau$ is the temperature in the softmax function}, $\varphi(z^t)$ is a linear transformation of $z^t$. The KL divergence of the posterior distribution of $Z^t$ and the prior distribution of $Z^g$ is then defined by:
\begin{equation}\small
    {\rm KL}(p(Z^t|x^t,\epsilon)|r(Z^g)) = \sum_{k=1}^{K} \eta^t_{k} \text{log}\frac{\eta^t_{k}}{\rho^g_{k}}
\label{eq:KL}
\end{equation}
\rev{where $\eta$ is the derived cluster-based distribution defined in Eq. (\ref{eq:qz_softmax}), $\rho$ is the prior distribution calculated from the user-item interaction graph, defined in Eq. (\ref{eq:cluster}).}

\paragraph{Reconstruction Loss.}
As discussed in Section \ref{sec:GIB}, the reconstruction term in Eq. (\ref{eq:obj_Lagrange}) has the property, $Q(x^t)  \varpropto {\rm log(exp}(-||\Bar{x}^t_{\epsilon} - x^t||^2))$. Therefore, it can be defined as the mean squared error function:
\begin{equation}\small
 Q(x^t)  = \mathbb{E}_{x^t \sim X^t}  \mathbb{E}_{\epsilon \sim p(\epsilon)} [||\Bar{x}^t_{\epsilon} - x^t||^2], 
  \label{eq:obj_square}
\end{equation}
\noindent where $x^t$ denotes the input user or the item representation, and $\Bar{x}^t_{\epsilon}$ denote the reconstructed representation. 

\paragraph{Rating Prediction and Final Objective Function.}
After obtaining the regularised latent textual representations for user $z^t_u$ and item $z^t_i$, we add their corresponding ID features to obtain the final user and item embeddings $\zeta_u$ and $\zeta_i$ for rating prediction, $\zeta_u = W_uz^t_u+\epsilon_{u}$, $\zeta_i = W_iz^t_i+\epsilon_{i}$, where the ID features $\epsilon_{u}$ and $\epsilon_{i}$ are generated by feeding the user ID and item ID to an embedding layer. Inspired by the latent factor model in recommendation systems, we introduce the global biases for users and items in the final prediction layer as $\hat{r}_{ui} = f_\text{cls}(\zeta_u,\zeta_i)+b_{u}+b_{i}$, where \yhq{$f_\text{cls}$ combines the user and item features into a scalar}, $b_{u}$ and $b_{i}$ are bias derived from $\epsilon_{u}$ and $\epsilon_{i}$. The regression loss $\mathcal{L}_{r}$ of the predicted rating is calculated as the MSE on the given user-item pair. Combining all the components above, we derive the training objective as follows:

\begin{equation}
    \mathcal{L} = \mathcal{L}_{r}+O^t_{IB} +  \mathcal{R}_{centroid} \label{eq:finalObjective}
\end{equation}

\paragraph{\rev{Time Complexity and Scalability Analysis}}
\rev{In this subsection, we conduct a complexity analysis of both \texttt{\textbf{GIANT}} and LightGCN, which is our chosen base graph model. Suppose the number of user nodes is $m$, the number of item nodes is $n$, the data sparsity is defined as $\alpha$, and edges in the user-item
interaction graph are $E=\alpha \cdot m \cdot n$. The average number of reviews for users and items are $|D_{u}|$ and $|D_{i}|$, respectively. Let $G$ denotes the number of GCN layers. $d$ is the graph node embedding size, as well as the dimension of cluster centroid vector $\mathcal{C}$. $K_u$ is the number of user clusters, $K_i$ is the number of item clusters, $T$ is the maximum number of iterations for K-Means~\footnote{Noted that we use the K-Means package within Scikit-Learn for performing clustering, and the average computational complexity is recorded in the table.}. For simplicity, all the kernels in CNN are assumed having a uniform size of  $c$, and there are a total of $f$ kernels. The padding length for each review text is denoted as $L$. }

\rev{We use the user and item embeddings obtained from well-trained LightGCN. Its complexity analysis results were reported in~\cite{wu2021self}. The additional computations involve several steps, including K-means for user/item clustering, final rating prediction using a VAE framework with rate prediction loss $\mathcal{L}_{r}$, information bottleneck regularisation $O^{t}_{IB}$, and the centroid regularisation $R_\text{centroid}$. For the textual feature extraction, all the reviews associated with a single user or item are aggregated into one, which is then fed into the CNN. This leads to an expected input channel size of $m|D_{u}|$ for users and $n|D_{i}|$ for items. Consequently, the expected total computational complexity can be expressed as $\mathcal{O}(m|D_{u}|Lc^{2}f)+\,\mathcal{O}(n|D_{i}|Lc^{2}f)$. For rate prediction, we only calculate the dot-product for all user-item interactions, resulting in a complexity of $\mathcal{O}(E)$. For the calculation of the KL divergence in Eq. (\ref{eq:KL}), the complexity is determined by $(K_u+K_i)$ for each user-item pair. Since the dimensions of the user and item latent variables are identical to $K_u$ and $K_i$, respectively, the reconstruction complexity for each $(u,i)$ pair is also $(K_u+K_i)$ . }  
\begin{table}[h]
    \centering
    \begin{tabular}{c|l|c}
\hline
\multirow{3}{*}{LightGCN} & Adjacency& $\mathcal{O}(|E|)$ \\
    & GCN & $\mathcal{O}(|E|^{2}Gd)$\\
    & BRP Loss & $\mathcal{O}(|E|d)$\\
    \hline
K-Means & clustering&$\mathcal{O}(K_{u}mT)$+$\mathcal{O}(K_{i}nT)$\\
\hline
CNN & Feature extractor & $\mathcal{O}(m|D_{u}|Lc^{2}f)+\,\mathcal{O}(n|D_{i}|Lc^{2}f)$ \\
\hline
   \multirow{3}{*}{VAE} &Rate Prediction & $\mathcal{O}(E)$\\
   & $O_{IB}$ & $\mathcal{O}(EK_u+EK_i)$
\\
   & $R_\text{centroid}$ & $\mathcal{O}(dEK_u+dEK_i)$\\
\hline
    \end{tabular}
    \caption{The analytical time complexity
of \texttt{\textbf{GIANT}}. }
    \label{tab:my_label}
\end{table}
\rev{Since the number of clusters is much smaller than the number of items or users, the expected increase in complexity primarily arises from the K-Means clustering, textual feature extraction, and rate prediction steps, roughly speaking:}
\begin{align}
  \mathcal{O}&((K_uT+|D_u|Lc^{2}f)m+(K_iT+|D_i|Lc^{2}f)n)+ \nonumber\\
  \mathcal{O}&((dK_u+K_u+dK_i+K_i+1)E)
\end{align}
\rev{In our highest configuration, we set $d$ to 64, $K_u$ and $K_i$ to $64$, $|D_u|$ and $|D_i|$ to $100$, $L$ to $32$, kernel size to $3$, and the number of kernels to $3$. Consequently, the dominant terms are the second coefficient for $m$ and $n$, approximately $86,400 \cdot m$ and $86,400 \cdot n$. This allows us to simplify the computational complexity as $\mathcal{O}(|D_u|Lc^{2}fm+|D_u|Lc^{2}fn+(dK_u+dK_i)E)$. Taking into account the sparsity of our datasets, as shown in Table~\ref{tab:dataset}, 
we can approximate $E$ as $(3.5e-3)mn$, 
the overall complexity is primarily determined by $|D_u|Lc^{2}fm+|D_i|Lc^{2}fn+14mn$. \textsl{This complexity scales linearly with the number of reviews for each user or item}.}

\rev{When comparing our cost to that of LightGCN, which is expressed as  $2|E|^{2}G$, where $G$ represents the number of GCN layers, we find that our time complexity becomes comparable to LightGCN only when the result of $|D_u|Lc^{2}fm+7mn$ approaches ${(\alpha m n)}^2$. 
In practical terms, in our experimental configurations, these two results are within the same level. 
Therefore, our time complexity is nearly equivalent to that of existing graph-based recommender systems.
}

\section{Sentence-level Explanation Generation}
\label{sec:gen_ex}
\hanqi{Our variational framework can identify the top-$k$ words for each latent dimension.} To make it easier for humans to understand the rationales behind model decisions, we extract the most relevant \emph{review sentences} (see the extracted explanations in Figure~\ref{fig:intro} and examples in Figure ~\ref{fig:casestudy}) from all the candidate reviews as interpretations for a specific user-item pair ($u$-$i$).

\yhq{In particular, our \texttt{\textbf{GIANT}} model infers the latent factors $z^t$. The most prominent dimension in the latent factor represents the assigned cluster for user $u$ or item $i$. Recall that the number of clusters derived in the graph space equals to the number of latent dimensions. Based on that, we can identify the user candidate reviews as all the past reviews on item $i$ written by the users assigned to the same cluster as $u$. Similarly, we identify item candidate reviews from the past reviews on the item $i$ which have their most prominent topic the same as the item latent topic.} \yhq{Afterwards, we represent each topic by its top-associated 5 words and derive the topic representation by the aggregated word-level GloVe~\footnote{https://nlp.stanford.edu/projects/glove/} word embeddings.~\footnote{We also experimented with more representative words but observed less discriminative topic clusters.} These words are selected based on the TFIDF on the reviews in the same topic cluster, with stop words filtered. The review sentence representations are also derived based on the aggregated constituent word GloVe embeddings. The most relevant sentences from user candidate reviews can then be extracted as a summary of the user $u$'s preferences based on their cosine similarity with the user latent topic representation. Similarly, the most relevant sentences from item candidate reviews are extracted as a summary of item $i$'s characteristics based on their cosine similarity with the item's latent topic representation.}

\section{Experiments Results on Recommendation}

\subsection{Experimental Setup}
\paragraph{Dataset and Data Processing} We choose three commonly used e-commerce datasets, including \textsl{BeerAdvocate}~\cite{DBLP:conf/icdm/McAuleyLJ12} and two amazon review datasets, \textsl{Digital Music} and \textsl{Office Products}~\cite{DBLP:conf/www/HeM16}, which not only contain the interaction interactions between user and item but also the review texts. Amazon review data~\footnote{\url{https://jmcauley.ucsd.edu/data/amazon/}} is one of the popular datasets, consisting of 24 kinds of products. As our method derive the user/item mainly based on their reviews, we use the dense subset, 5-core, of the amazon review dataset that extracts the user-item pairs, such that each of the remaining users and items have at least 5 reviews each. 
Table~\ref{tab:dataset} summarizes the statistics of the three datasets. \rev{In order to enhance the performance of our recommender system, we employ a filtering mechanism for entries (user/item) associated with either reviews of significant length or an exceptionally large number of reviews. This is done because both the length and the number of reviews follow a long-tail distribution, as described in \cite{DBLP:conf/www/ChenZLM18}.} Specifically, we filter out users or items with over 100 reviews/interactions and truncate the reviews to  lengths of [200,400,300] for the three datasets, respectively. Additionally, we randomly select 80\%, 10\%, 10\% as the train/validation/test sets.

\paragraph{Evaluation Metrics} We use Root Mean Squared Error (RMSE) and Mean Absolute Error (MAE) to evaluate the rating prediction accuracy. We also report the ranking-based evaluation results, i.e., Precision and Recall to measure the overlapping between the recommended items and the target items.
\renewcommand{\arraystretch}{1.3}
\begin{table}[ht]
    \centering
    \resizebox{0.48\textwidth}{!}{%
    \begin{tabular}{lcccccc}
    \toprule[1pt]
        Datasets & \#Train & \#users & \#items&AveRate&Sparsity  \\
        \hline
      \textsl{BeerAdvocate}  &35450 & 6939&13122&3.74&4.4e-4 \\
      \textsl{Digital Music} &51768&5541&3568&4.22&2.9e-3\\
      \textsl{Office Products} &35999&4902&2364&4.33&3.5e-3\\
\bottomrule[1pt]
    \end{tabular}
    }
    \caption{All the datasets have ratings in the range of 1-5. The BEER dataset has the smallest sparsity.}
    \label{tab:dataset}
\end{table}
\paragraph{\textbf{Baselines}}
We compare with several open-source recommenders, 
including \textsl{HFT}~\cite{10.1145/2507157.2507163}, \textsl{DeepCoNN}~\cite{DBLP:conf/wsdm/ZhengNY17}, and \textsl{NARRE}~\cite{DBLP:conf/www/ChenZLM18}. \textsl{HFT} combines reviews with ratings and uses an exponential transformation function to link review text and the ratings. \textsl{DeepCoNN} uses a shared layer for interaction modeling the users and items, which is on top of the two encoders for the users and items, respectively. \textsl{NARRE} is an interpretable recommender, which introduces review-level attentions to select important reviews and incorporates the user and item IDs as discriminative features for rating prediction and uses the text spans with higher attention as recommender rationales. Our \texttt{\textbf{GIANT}} is also built on the \textsl{NARRE}, but with a proposed geometric prior in an encoder-decoder structure. \rev{More recently, a graph model called \textsl{RGCL}~\cite{10.1145/3477495.3531927} was introduced, which incorporates a contrastive loss and takes into account both user-item interactions and review semantics in its node features. Our model similarly emphasises the significance of user-item interactions in review-based recommender systems, but it introduces a novel regularisation technique based on clustering.} Furthermore, we compare with several other encoder-decoder structures with different regularizes on top of \textsl{NARRE} to learn the latent variable as a baseline to highlight the difference with our proposed regularisation, i.e., \textsl{AutoEncoder}, \textsl{StandPrior} and \textsl{WassersteinVAE} and \textsl{IndivPrior}:

\begin{itemize}[leftmargin=*]
\item \noindent\textsl{\underline{StandPrior}} (aka. \textsl{StandVAE}) uses two VAEs for users and items respectively, each with the standard Gaussian distribution $\mathcal{N}(0,1)$ as the prior. 
\item \noindent\textsl{\underline{WassersteinVAE}} differs from the \textsl{StandPrior} in using the Wasserstein metric~\cite{DBLP:journals/corr/abs-1711-01558} to calculate the distribution discrepancy between posterior $\eta$ and prior $\rho$. 
\item \textsl{\underline{IndivPrior}} assigns a separate Gaussian prior to each user or item with its mean value calculated from the user's or item's corresponding representation derived from the user-item interaction graph, i.e., 
$p(z^t_u)=\mathcal{N}(W_{u}\xi_{u},\bm{I})$, $p(z^t_i)=\mathcal{N}(W_{i}\xi_{i},\bm{I})$, where $\xi_{u}\in\mathbb{R}^{d}$, $\xi_{i}\in\mathbb{R}^{d}$ denote the user and item graph node, $W_{u}$ and $W_{i}$ are learnable parameters in a linear layer of size $(d \times d)$, and $d$ is the dimension of graph node embeddings. 
\end{itemize}

\paragraph{Training Procedure}
We train the model by minimising the objective function defined in Eq. (\ref{eq:finalObjective}). \hq{The $\beta$ for the KL divergence term is set to 0.01 and the $L_{2}$ regularisation is used for all the model parameters with the weight of 0.001.}
We only train the prior-centralisation term for 0.5 proportion to approximate the cluster centroids for the later training. \hq{The KL term in $O_{IB}$ should be introduced later until the encoder centroid is well trained. To do so, we follow a similar cyclical schedule~\cite{DBLP:conf/naacl/FuLLGCC19} to gradually adjust the 
anneal factor $\lambda$ in each epoch. Specifically, we first train the model without the KL term for 0.5 proportion, then anneal it from 0.5 to 1 for 0.25 proportion, and finally fix $\lambda = 1$.~\footnote{Hyper-parameter settings are included in Appendix A.}}

\subsection{Rating Prediction Results}

The rating prediction results of our model in comparison with baselines on all datasets are given in Table~\ref{tab:main_results}. We have the following observations. (1) \textsl{DeepCoNN} built on the stack of non-linear neural networks for review semantic modeling outperforms \textsl{HFT} which leverages an exponential transformation function to link topic distributions in review text and latent factors derived from ratings. This shows the superiority of deep learning for feature extraction. (2) \textsl{AutoEncoder} gives slightly better performance than \textsl{NARRE} in \textsl{BeerAdvocate}, which shows its effectiveness of extracting key contextual information for rating prediction. (3) \rev{\textsl{RGCL} emerges as the leading baseline model and even manages to achieve  competitive results on both the \textsl{Beer} and \textsl{Music} datasets when compared to our \textsl{GIANT}. However, when considering the Office dataset, its overall performance falls short of ours. The average RMSE/MAE values across the three datasets are 84.49/63.49 for \textsl{RGCL} and 84.18/62.87 for\textsl{GIANT}, indicating that our model outperforms \textsl{RGCL}. 
} The results demonstrate the effectiveness and robustness of our proposed information bottleneck regularisation applied on the latent semantic space. By doing this, the users and items can be grouped into different clusters according to the interaction data, which is not the case in other baseline recommenders.

\paragraph{Effect of our proposed information bottleneck regularisation}
We then study the effects of different information bottleneck, i.e., \textsl{StandPrior}, \textsl{WassersteinVAE} and \textsl{IndivPrior}. From the results in the last group of Table~\ref{tab:main_results}. It can be seen that \textsl{WassersteinVAE} and \textsl{StandPrior} demonstrate overall better results than \textsl{IndivPrior} and all the baselines in Table~\ref{tab:main_results}. The improvement can be explained by the nature of VAE, which is particularly beneficial when dealing with sparse data where few observations are available. However, \textsl{IndivPrior}, which imposes a separate Gaussian prior on individual user/item, shows worse performance compared to the other three variational frameworks, and is even worse than \textsl{AutoEncoder}. \hq{Our model benefits from a rich prior derived from user/item clusters, while avoiding using a global normal prior (\textsl{StandPrior} and \textsl{WassersteinVAE}) or a separate instance-level prior for each user/item (\textsl{IndivPrior}).}

\begin{table}[h]
\centering
\resizebox{0.48\textwidth}{!}{%
\begin{tabular}{l|cc|cc|cc}
\toprule[1pt]
\multirow{2}{*}{\textbf{Models}}&\multicolumn{2}{c|}{\textsl{\textbf{BeerAdvocate}}} & \multicolumn{2}{c|}{\textsl{\textbf{Digital Music}}}&\multicolumn{2}{c}{\textsl{\textbf{Office Products}}}\\
\cmidrule{2-3}\cmidrule{4-5}\cmidrule{6-7}
& RMSE($\downarrow$)&MAE($\downarrow$)&RMSE($\downarrow$)&MAE($\downarrow$)&RMSE($\downarrow$)&MAE($\downarrow$)\\
\textsl{\textbf{HFT}}&79.81 &62.42 & 96.42& 74.76 &89.46  &68.57 \\
\textsl{\textbf{DeepCoNN}}&77.28&59.45&94.69&71.03&85.14&64.82\\
\textsl{\textbf{NARRE}}& 
76.80&58.94&\underline{93.69}&69.30&\underline{84.40}&63.40 \\
\textbf{\textsl{RGCL}} &\textbf{74.65}&\textbf{57.12}&\textbf{92.02}&\textbf{68.49}&86.81&64.87 \\
\midrule
\textsl{\textbf{AutoEncoder}}& 75.94 & \underline{58.93} & 93.97&\underline{69.13}&85.03&\underline{64.32}\\
\textsl{\textbf{StandPrior}}& 75.80 &58.73 & \underline{93.60}&\underline{69.21}&84.69&63.97\\
\textsl{\textbf{WassVAE}}& \underline{75.62} & \underline{58.06} & 93.81&69.32&\underline{84.20}&\underline{63.43}\\
\textsl{\textbf{IndivPrior}}& 76.43 &59.88  & 94.57&70.51&85.33&65.05\\
\midrule
\textsl{\textbf{GIANT}}&$\underline{75.36}^{*}$&$\underline{57.87}^{*}$&$\underline{92.87}^{*}$&$\underline{68.68}^{*}$&$\bm{84.32}$&$\bm{62.05}^{*}$\\
\bottomrule[1pt]
    \end{tabular}
    }
    \caption{Performance comparison in RMSE (\%) and MAE (\%) for all methods. 
    * denotes the statistical significance for 
    $p < 0.01$, 
    compared to the best baseline. The best results throughout all the rows are in bold, the best results in each group are underlined. }
    \label{tab:main_results}
\end{table}

\paragraph{Ranking-based Results}
\yhq{In addition to the accuracy-based metrics, i.e., MSE and MAE, we also evaluate models using the ranking-based metrics, i.e., \rev{HitRatio$@$10} and \rev{NDCG$@$10} for the two best baselines (Table~\ref{tab:rank-based results}).  \rev{Without evaluation on the whole item list, we followed the common practice in~\cite{10.1145/3038912.3052569,DBLP:conf/www/ElkahkySH15} that randomly samples negative 100 items that are not interacted by the user, ranking the test item among the 100 items. Then the ranked list is truncated at 10 for both HitRatio@10 and NDCG@10.}}
\rev{The performance trends across different datasets diverge from those observed in accuracy-based metrics. \textsl{RGCL} demonstrates comparable performance on the Office dataset, whereas its  performance on the \textsl{BeerAdvocate} dataset is much worse, achieving a HitRatio of 0.47 compared to 0.62 for \textsl{\textbf{GIANT}}. However, our \textsl{GIANT} still maintains an overall  superior performance in rank-based metrics. Specifically, on the HitRatio$@$10 
metric, our model achieves an score of 0.65, 
while \textsl{RGCL} achieves an score of 0.61. 
}
\textsl{WasersteinVAE} appears to be the second-best encoder-decoder model (the second group). The relative rank-based performances across the two groups in Table~\ref{tab:main_results} are different, while the relative performances within encoder-decoder structured models are similar to accuracy-based metrics, i.e., MSE and \texttt{\textbf{GIANT}} in general achieves the best results on the two kinds of metrics.

\begin{table}[h]
\centering
\resizebox{0.48\textwidth}{!}{%
\begin{tabular}{l|cc|cc|cc}
\hline
\multirow{2}{*}{Models}&\multicolumn{2}{c|}{\textsl{\textbf{BeerAdvocate}}}&\multicolumn{2}{c|}{\textsl{\textbf{Digital Music}}}&\multicolumn{2}{c}{\textbf{\textsl{Office Products}}}\\
\cmidrule{2-7}
& HitRatio & NDCG& HitRatio & NDCG & HitRatio& NDCG \\
\hline
\textsl{\textbf{PureMF}}& 0.43 &0.23 &0.78&0.52&0.61&0.33 \\
\textsl{\textbf{SVD++}}& 0.41 & 0.21 &0.52&	0.28&0.55&0.28\\
\textsl{\textbf{NeuMF}}& 0.41 &	0.21 & 0.78 & 0.52 & 0.63 &	0.35\\
\textsl{\textbf{NARRE}}& 0.61 & 0.52 &0.65&0.59&0.59&0.64\\
\textsl{\textbf{RGCL}}&0.47&0.55 &\textbf{ 0.72} &0.63&\textbf{0.64}&\textbf{0.72}\\
\midrule
\textsl{\textbf{AutoEnc}}& 0.52&0.49&0.62&0.60&0.61&0.62\\
\textsl{\textbf{StandPrior}}&0.55&0.54&0.68&0.63&0.61&0.62\\
\textsl{\textbf{WassVAE}}& 0.55 & 0.54& 0.69&0.65&0.61&0.60\\
\textsl{\textbf{GIANT}}& \textbf{0.62}&\textbf{0.57}&0.69&\textbf{0.67}&\textbf{0.64}&0.64\\
\midrule
    \end{tabular}
}
    \caption{\rev{The evaluation results based on ranking, namely,  HitRatio$@$10 and \rev{NDCG$@$10} for all the baseline and our proposed \texttt{\textbf{GIANT}}.}}
    \label{tab:rank-based results}
\end{table}

\subsection{Contribution from Various Loss Terms}
We study the effects of the three loss terms in rating prediction accuracy and \emph{Diversity} of the latent variables. As variational networks could easily collapse into an unconditional generative model, i.e., in the extreme case, all the input will be mapped into the same latent code~\cite{DBLP:conf/iclr/MaZH19}. We use the dimension index whose corresponding latent value is the maximum as the cluster assignment ID. We then derive the cluster assignment results $\mathcal{A} \in \mathbb{R}^{N\times K}$, where $N$ is the number of test samples, $K$ is the cluster number.
The diversity is calculated based on entropy $H(X) = -\sum_{k \in K} p(k) \log(p(k)))$, where $p(k)$ is the fraction of the number of samples falling into the $k$-th cluster among all the samples. A larger value means better diversity. We record the largest diversity between user and item latent variables. The results are shown in Table \ref{tab:ablationstudy}.

\begin{table}[h]
\centering
\resizebox{0.49\textwidth}{!}{%
\begin{tabular}{l|lll|lll|lll}
\toprule[1pt]
\multirow{2}{*}{\textbf{Variants}}&\multicolumn{3}{l|}{\textsl{\textbf{BeerAdvocate}}}&\multicolumn{3}{l|}{\textsl{\textbf{Digital Music}}}&\multicolumn{3}{l}{\textsl{\textbf{Office Products}}}\\
\cmidrule{2-4}\cmidrule{5-7}\cmidrule{8-10}
 & \rev{NDCG} & RMSE &Div & \rev{NDCG}& RMSE&Div&  \rev{NDCG}& RMSE&Div\\
\midrule
\textbf{Full Model} &0.57&75.36&1.89&0.67&92.87&3.46&0.64&84.32&1.58 \\
\textbf{-w/o}$\bm{\mathcal{R}_{cent}}$ &0.55 &75.96&0.04&0.65&92.62&0.07&0.62&84.58&0.02 \\
\textbf{-w/o KL term} &0.55&75.64&1.43&0.65&93.83&2.86&0.61&84.23&1.21 \\
\textbf{-w/o $\bm{Q(x^t)}$ } &0.52&76.02&1.04&0.61&93.95&3.10&0.59&84.94&0.93 \\
\midrule
$\bm{Q(x^t)}$ \textbf{Cosine } &0.56&75.35&0.11&0.67&92.85&1.76&0.63&84.97&0.07 \\
\bottomrule[1pt]
\end{tabular}
}
\caption{\rev{Rating prediction results, including metrics such as \emph{NDCG$@10$}}, \emph{RMSE}(\%) and \emph{Diversity} across model variants. A lower diversity indicates that the learnt latent variables for different samples tend to collapse into a similar latent code.}
\label{tab:ablationstudy}
\end{table}

We observe that 
the removal of $Q(x^t)$ leads to the largest performance degradation. After removing $\mathcal{R}_{centroid}$, the latent variable diversity shrinks to near zero, indicating nearly all the latent variables fall into the same cluster. This highlights the capability of our prior-centralisation term in
enabling the latent variables $z^t$ effectively reflect the soft cluster assignments. We also found that replacing the MSE in $Q(x^t)$ with the cosine similarity ($1-cosine$) reduces the latent variable diversity significantly. This can be partly explained by the fact that the cosine similarity focuses on measuring the angle between the input $x^t$ and the reconstructed output $\Bar{x}^t$, ignoring the magnitude of the vectors which is however important in our case. 


\section{Latent Variable and Interpretability Evaluation}
\hanqi{In addition to the rating prediction, we design experiments to evaluate the generated latent variables and explore the improvement source. Firstly, we evaluate the generated latent clusters in terms of semantic coherence and separability. Then, we examine the effects of the identified important latent dimensions on rating prediction. Finally, we conduct a human evaluation to study the interpretability.}
\subsection{Cluster Coherence and Separability}
\label{sec:topic_extract} 
\paragraph{Semantic Coherence} To verify the capability of creating \emph{semantically coherent clusters}, we propose to measure cluster coherence as the average cosine similarity between every review document pair within a cluster. We first obtain the document-level review representations by feeding the reviews from the test set to our pre-trained CNN encoder. 
We then calculate the average cosine similarity 
between the representations of each review pair 
in a cluster, and are further averaged across all clusters (Table~\ref{tab:coherent-cluster}). We use the clusters generated via applying KMeans on learnt graph node embedding as a comparison, i.e., $\rho$ in Eq. (\ref{eq:cluster}).

\renewcommand{\arraystretch}{1.3}
\begin{table}[h]
\centering
\resizebox{0.48\textwidth}{!}{%
\begin{tabular}{b{0.08\textwidth}b{0.045\textwidth}b{0.045\textwidth}b{0.045\textwidth}b{0.045\textwidth}b{0.045\textwidth}b{0.045\textwidth}}
\toprule[1pt]
\multirow{2}{*}{\textsl{\textbf{Models}}}&\multicolumn{2}{c}{\textbf{\textsl{BeerAdvocate}}} & \multicolumn{2}{c}{\textsl{\textbf{Digital Music}}}&\multicolumn{2}{c}{\textsl{\textbf{Office Products}}}\\
\cmidrule(lr){2-3}\cmidrule(lr){4-5}\cmidrule(lr){6-7}
& User&Item&User&Item&User&Item\\
\hline
\textsl{\textbf{Graph}}&0.092& 0.085& 0.155& 0.214& 0.245&0.200 \\
\textsl{\textbf{GAINT}}&0.490&0.481&0.457&0.471&0.492&0.407\\
\bottomrule[1pt]
\end{tabular}}
\caption{A larger similarity value means a better coherence. \texttt{\textbf{GIANT}} generates significantly more coherent clusters than graph clusters.}
\label{tab:coherent-cluster}
\end{table} 

\paragraph{Cluster Separability} As each dimension of the latent variables $z^t$ corresponds to a cluster, for the $k$-th dimension, we can search for its relevant reviews which have the highest value in $z^t_{k}$ and list the most frequent words in the review set as the representative topic words~\footnote{We exclude the stopwords and most frequent words appeared in all clusters.} Representative words in randomly selected three clusters on the three datasets are shown in Table~\ref{tab:our_clusterwords}.  For our model, the most prominent words are different across different clusters (thus being coloured), i.e., `\emph{brown roasted}', `\emph{pine, caramel}' and `\emph{good flavour}'. We identify `\emph{ABBA}' and their popular songs, computer peripherals such as `\emph{network cable}' and `\emph{router}' in \textsl{Digital Music} and \textsl{Office Products} datasets, respectively.

\yhq{We find it hard to see clear topic separations from the \textsl{WassersteinVAE} results. For example, Cluster 2 and 3 share the same 3 words, `\emph{sweet}', `\emph{light}' and `\emph{malt}' out of their top 4 words. In addition, `\emph{carbonation}' in Cluster 2 and `\emph{whitehead}' in Cluster 3 are both related to beer foam. The results show that with the incorporation of the priors derived from user/item clusters, our proposed approach is able to learn latent variables in the review semantic space which can produce better separable topic clusters.}  

\renewcommand{\arraystretch}{1.3}
\begin{table*}[t]
\centering
\resizebox{0.95\textwidth}{!}{%
\begin{tabular}{p{0.2cm}p{8.5cm}p{8.5cm}}
\toprule[2pt]
&\texttt{\textbf{GIANT}} & \textsl{\textbf{WassersteinVAE}} \\
\hline
\multirow{3}{*}{\rotatebox[origin=c]{90}{\textsl{Beer }}}& \textcolor{black}{\textbf{pour}}, color, \textcolor{black}{\textbf{brown}}, feel, \textcolor{black}{\textbf{roasted}}, malts, almost, moderate, coffee, dark&light,
  color,
  malt,
  \textcolor{black}{\textbf{glass}},
  poured,
 \textcolor{black}{\textbf{ drink}},
  sour,
  pour,
  mouthfeel,
 \textcolor{black}{\textbf{alcohol}}\\
 & \textcolor{black}{\textbf{good}}, \textcolor{black}{\textbf{flavour}}, dark, \textcolor{black}{\textbf{better}}, brew, style, bad, Canadian, nose, drinking&aroma,
  light,
  sweet,
  malt,
  \textcolor{black}{\textbf{white head}},
  mouthfeel,
  \textcolor{black}{\textbf{nose}},
  hint,
  \textcolor{black}{\textbf{brew}},
note\\
 & \textcolor{black}{\textbf{pine}}, \textcolor{black}{\textbf{caramel}}, lacing, \textcolor{black}{\textbf{citrus}}, mouthfeel, hint, ipa, note, body, strong&sweet,
  \textcolor{black}{\textbf{coffee}},
  light,
  malt,
  aroma,
  \textcolor{black}{\textbf{carbonation}},
  pour,
  mouthfeel,
  hint,
  color\\
\hline
\multirow{6}{*}{\rotatebox[origin=c]{90}{\textsl{Music}}}& \textcolor{black}{\textbf{dr dre}}, \textcolor{black}{\textbf{prince}}, \textcolor{black}{\textbf{used known}}, west coast, mirrors, love, hate, westside story, jadekiss, always&quot,
  album,
  \textcolor{black}{\textbf{Ask Rufus}},
  song,
  \textcolor{black}{\textbf{funk}},
  \textcolor{black}{\textbf{voice}},
  music,
  band,
  soul,
  love
  \\
 \cline{2-3}
&\textcolor{black}{\textbf{bangles}}, \textcolor{black}{\textbf{vicki}}, \textcolor{black}{\textbf{vocal}}, place, live, hero, takes, fall, liverpool, beatles&
song,
  \textcolor{black}{\textbf{everglow}},
  album,
  band,
  \textcolor{black}{\textbf{beautiful}},
  track,
  \textcolor{black}{\textbf{lyrics}},
suspension,
  record,
Christian,
 \\
 \cline{2-3}
& \textcolor{black}{\textbf{dancing queen}}, \textcolor{black}{\textbf{take chance}}, \textcolor{black}{\textbf{mamma mia}}, abba, greatest hit, money money, gimme gimme, super trouper, knowing knowing, gold&
record,
album,
  song,
  band,
  track,
  time,
  sound,
  \textcolor{black}{\textbf{head heart}},
  \textcolor{black}{\textbf{folk}},
  \textcolor{black}{\textbf{live}},
love
\\
\hline
\multirow{5}{*}{\rotatebox[origin=c]{90}{\textsl{Office}}}&\textcolor{black}{\textbf{set}} , \textcolor{black}{\textbf{computer}},  \textcolor{black}{\textbf{wireless}}, network, cable, router, edit, pencil, ink&\textcolor{black}{\textbf{scan}},
  \textcolor{black}{\textbf{software}},
  pencil,
  \textcolor{black}{\textbf{Mac OS}},
  work,
  printer,
small,
  printed,
  color,
  file
\\
 \cline{2-3}
&\textcolor{black}{\textbf{photo}}, \textcolor{black}{\textbf{desk}}, \textcolor{black}{\textbf{laser}}, scanner, work, scan, epson, scaning, enough, easy& easy use,
  \textcolor{black}{\textbf{dry erase}},
  folder,
  \textcolor{black}{\textbf{mouse pad}},
office,
  \textcolor{black}{\textbf{ink cartridge}},
  works,
  used,
post note\\
 \cline{2-3}
& pencil, \textcolor{black}{\textbf{tape}}, \textcolor{black}{\textbf{ink}}, pen, anyway, pen, \textcolor{black}{\textbf{rubber}}, clip, pretty, eraser&
\textcolor{black}{\textbf{recommend}},
  printer,
  folder,
  \textcolor{black}{\textbf{paper}},
  easy use,
  \textcolor{black}{\textbf{boxes}},
  feature,
  stapler,
  pencil,
  canon,
  color\\
\bottomrule[2pt]
    \end{tabular}}
    \caption{\yhq{The most prominent words (sorted by occurrence frequency) in three randomly selected clusters from \texttt{\textbf{GIANT}} and \textsl{WassersteinVAE} on \textsl{Digital Music} and \textsl{Office Products} datasets. For each dataset, we highlight the top 3 words that are not found in the other two clusters. \texttt{\textbf{GIANT}} generates better separable topics while \textsl{WassersteinVAE} fail to generate clear topic pattern.}}
    \label{tab:our_clusterwords}
\vspace{-5pt}
\end{table*}

\subsection{Comprehensiveness Evaluation by Perturbing on Latent Variables}
\begin{figure*}[t]
    \centering
    \includegraphics[width=0.80\textwidth]{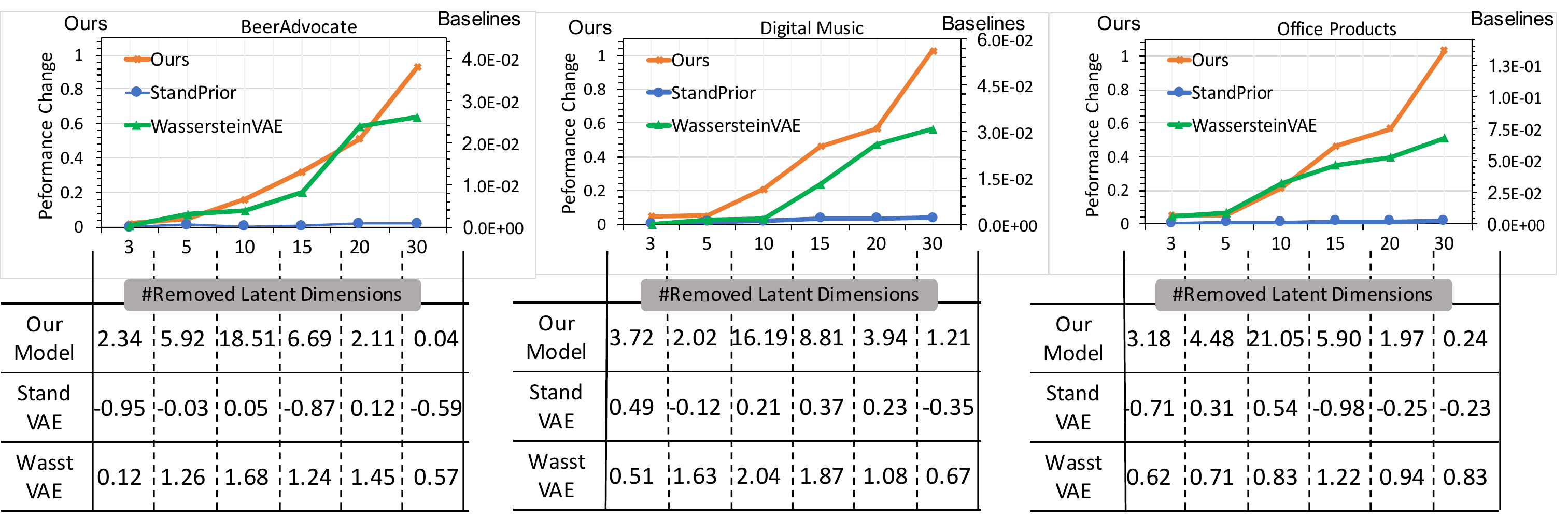}
    \caption{\textbf{Top}: The \textsl{Comprehensiveness} values by removing the top $k$
most important latent dimensions identified, $k \in \{3, 5, 10, 15, 20,30\}$. Ous model uses the left y-axis, the two baselines use the right axis. \textbf{Bottom Table}: Relative performance changes after subtracting the changes caused by random removal.}
    \label{fig:comprehensiveness}
\vspace{-15pt}
\end{figure*}

To explore the importance of our identified clusters, i.e., the latent dimension with a larger value, we calculate the performance change before and after removing the specific dimension and define \emph{Comprehensiveness} as: $\frac{\sum_{i}^{N}\left(r(z^i)-r(z^{i}_{\backslash{k}})\right)^{2}}{N}
$, Where $r(\cdot)$ is the predicted rating, $z^i$ is the latent variable for $i$-th evaluated user-item pair, $z^i_{\backslash{k}}$ is to remove the identified top $k$ latent dimensions from $z^i$. To this end, we replace the values in the top-$k$ dimensions with the average value of the latent variables according to~\cite{DBLP:journals/corr/FongV17}. In the Figure~\ref{fig:comprehensiveness} line chart, our model demonstrates a larger performance change as more important dimensions are removed.
We further randomly delete $k$ latent dimensions and calculate the relative performance change by subtracting the changes caused by random removal (Table below Figure~\ref{fig:comprehensiveness}). The relative changes in our model are most predominant, \yhq{followed by \textsl{WassersteinVAE}}, while relative changes of \textsl{StandPrior} are negative, showing that the random removal of latent dimensions can bring even larger changes to the predominant ones.

\begin{figure}[t]
\includegraphics[trim={2 6 8 0},clip,width=0.43\textwidth]{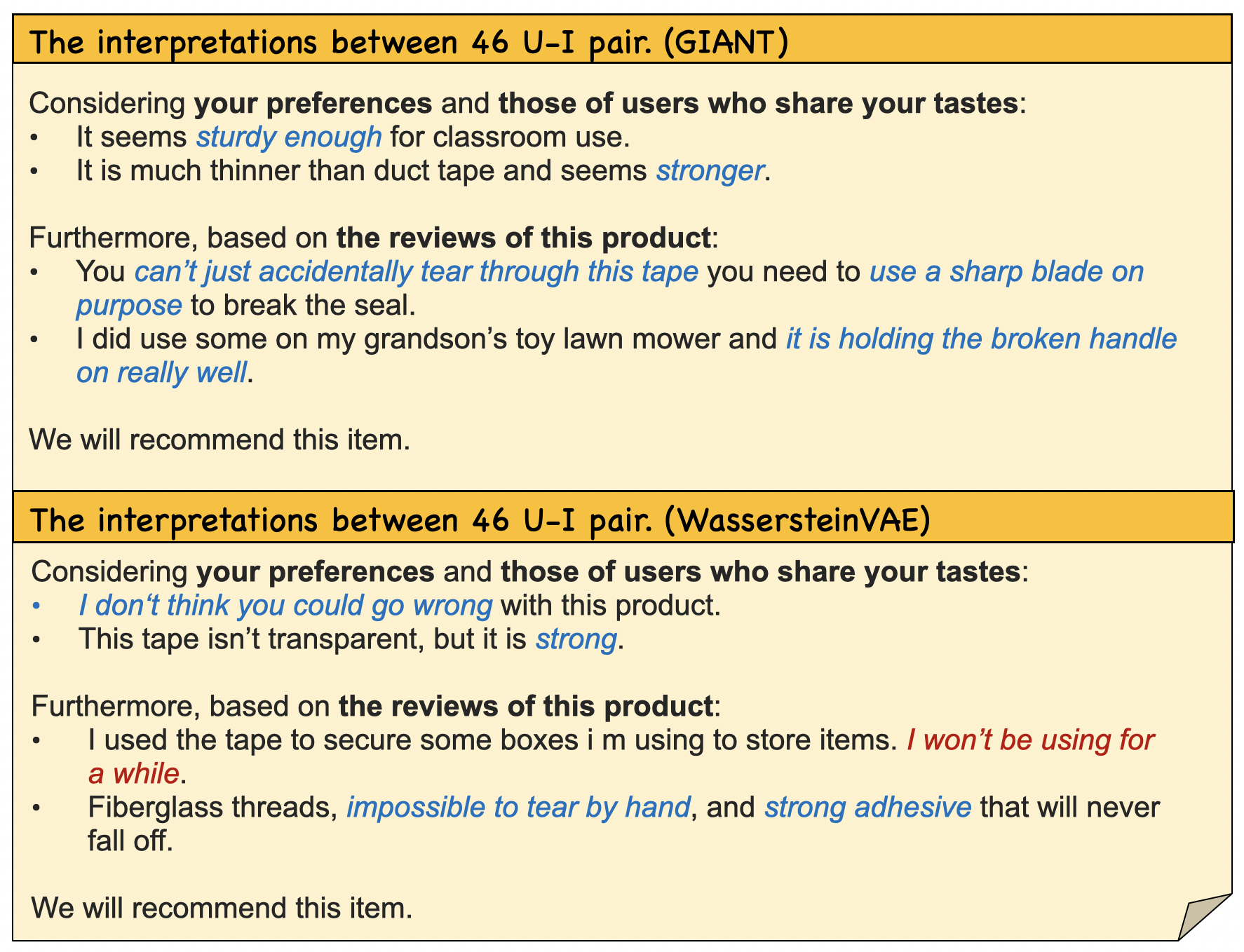}
    \caption{\yhq{Extracted interpretations from \textbf{\texttt{GIANT}} (upper) and \textsl{WassersteinVAE} (bottom) for the item, \emph{a tape}. In the example generated by \textbf{\texttt{GIANT}}, both the reviews/rationales from the user and item side focus on the \emph{`sturdy and strong'}, and reveal lots of important aspects for the recommendation. While the user reviews extracted by \textsl{WassersteinVAE}, e.g., \emph{`go wrong'} fail to capture the key factors, and the extracted item reviews are even contradictory, i.e., \emph{`won't be using'} to the recommended behaviour.
    }}
    \label{fig:casestudy}
\vspace{-5pt}
\end{figure}

\subsection{Human Evaluation of Generated Interpretations}
\label{subsec:casestudy}
We conduct a human evaluation to validate the
interpretability of our proposed method. To make it easier for humans to understand the rationales behind model decisions, we extract the most relevant sentences from user/item reviews as interpretations for a specific user-item pair (user $u$ and item $i$). Examples of generated interpretations are illustrated in Figure~\ref{fig:casestudy}, as well as the model's recommendation suggestion as \emph{will recommend} if the model's predicted rating score is above the average predicted rating; or \emph{won’t recommend}, otherwise. 

 We propose three evaluation metrics, \emph{relevance}, \emph{faithfulness} and \emph{informativeness} and ask three English-proficient human evaluators to give 1-5 score to the generated interpretations from \textsl{WassersteinVAE}, \textsl{NARRE} and \textsl{GIANT} on 120 randomly selected user-item pairs from each of the three datasets. The model identifiers are unknown to the evaluators.
 \begin{itemize}
 \item \emph{Relevance}. Do the extracted sentences from user reviews and item reviews pertain to the relevant topic, aspect, or subject? An interpretation with greater overlap in aspects should receive a higher score. 
\item \emph{Faithfulness}. Do the sentences extracted from user reviews and item reviews influence the recommendation suggestions made by the model? \rev{You should consider both the sentiment and the aspects described. For example, if both the selected reviews from the user and the item express positive sentiments regarding the ``\textit{prices}'' of the product, the faithfulness score should be rated as 5. Any discrepancies between the sentiment and aspects mentioned can lower this score. 
}
\item \emph{Informativeness}. Do the provided interpretations effectively reflect the user preferences and item characteristics? \rev{This criterion should remain separate from the previous ones, only focusing on the extent to which various aspects and their associated sentiments are discussed in the extracted reviews. A lower level of coverage of aspects results in a reduced \emph{Informativeness} score, as exemplified by a statement like, ``I really like this product!''}. 
\end{itemize}

Our human evaluation results presented in Table~\ref{tab:my_label} shows that both \textsl{WassersteinVAE} and \texttt{\textbf{GIANT}} are better in \emph{relevance}, which can be partly explained that they can extract sentences from reviews of like-minded users from the same user clustering results when the current reviews don't contain any key information. The superior performance over \textsl{WassersteinVAE} further shows the efficiency of our proposed geometric prior.
\begin{table}[ht]
\centering
\resizebox{0.48\textwidth}{!}{%
    \begin{tabular}{llccc}
\toprule[1pt]
\multicolumn{2}{l}{\textbf{Metric}} & {\textbf{Relevance}} &{\textbf{Faithfulness}} & {\textbf{Informativeness}}\\
\hline
\multirow{3}{*}{\rotatebox[origin=c]{90}{\textsl{Beer}}}
 &
 \textsl{NARRE} & 2.7&	2.4	&3.1\\
\cline{2-5}
 &\textsl{WasserVAE} &\textbf{3.6} & 3.5&\textbf{3.5}\\
 \cline{2-5}
  &\textsl{GIANT} & 3.5&	\textbf{3.7} & \textbf{3.5}\\
  \hline
\multirow{3}{*}{\rotatebox[origin=c]{90}{\textsl{Music}}}
   &
   \textsl{NARRE}&4.4&	3.2	&3.8 \\
   \cline{2-5}
   &\textsl{WasserVAE} &4.0 & 3.3&3.0\\
    \cline{2-5}
  &\textsl{GIANT} & \textbf{4.7}&	\textbf{3.8} & \textbf{4.1}\\
  \hline
\multirow{3}{*}{\rotatebox[origin=c]{90}{\textsl{Office}}}
   &
   \textsl{NARRE}&2.7&2.8&3.9\\
   \cline{2-5}
   &\textsl{WasserVAE} &3.5 & 2.8&3.8\\
    \cline{2-5}
  &\textsl{GIANT} & \textbf{3.8}&	\textbf{3.3} & \textbf{4.2}\\
 \bottomrule[1pt]
    \end{tabular}
}
    \caption{Human evaluation results on \emph{relevance}, \emph{faithfulness} and \emph{informativeness} of the generated interpretations, which were assessed for 120 randomly selected user-item pairs from the three datasets. \rev{The Cohen’s Kappa co-
efficient among each pair of annotators for
the three metrics are 0.42, 0.38, and 0.47, respectively.} }
    \label{tab:my_label}
\vspace{-10pt}
\end{table}


\section{Conclusion}
\hq{In this paper, we leverage the user/item clusters sharing common interests/characteristics obtained from the user-item interaction graph to refine the review text latent factors via our proposed geometric information bottleneck (\texttt{\textbf{GIANT}}). 
We empirically show that \texttt{\textbf{GIANT}} is better in learning a semantically coherent and interpretable latent space and the generated explanations are more faithful to the model decisions, while achieving comparable rating prediction accuracy on three commonly used datasets.}

\section{Acknowledgement}
This work was funded by the UK Engineering and Physical Sciences Research Council (grant no. EP/T017112/1, EP/V048597/1, EP/X019063/1). HY receives the PhD scholarship funded jointly by the University
of Warwick and the Chinese Scholarship Council. YH is supported by a Turing AI Fellowship funded by the UK Research and Innovation (grant no. EP/V020579/1).

\bibliography{icml2023}
\bibliographystyle{IEEEtran}
\newpage
\appendices
\section{Hyper-Parameter Settings}
\label{sec:hyper-parameter}
We use Adam for optimization with the initial learning rate set to 0.001. The batch size is set to 32, 64, and 32, and training epochs are 10, 10, and 15 for the three datasets, respectively. 
 The adjacency matrix in the graph is binary: user-item entries with interacted ratings higher than the average corpus-wide rating is set to 1, and 0 otherwise. We stop training GCN when it reaches the best recall value in the validation set. The cluster centroid vectors are derived by K-means. The graph node feature dimension $d$ is 64, and the number of clusters \textsl{K} are searched in $[16,32,64,128,256,512]$. In CNN review encoding, the word embedding and ID embedding are both initialized by the uniform distribution $(-0.1,0.1)$, their dimensions are set to 300 and 64, respectively. We use two CNN kernels with size 2 and 3, and the dimension of 32 to encode the reviews. As such, the CNN-encoder outputs the review representations with dimension of 64 (i.e., 32$\times$2). To combine multiple reviews for each user or item, we use the attention mechanism where the attention weights are derived through two consecutive linear transformation layers of $64\rightarrow32\rightarrow1$.
In \texttt{{GIANT}}, the latent variable dimension is set to the same as the number of clusters $K$. The softmax temperature $\tau$ is searched in $[1,2,4,6,8]$. The non-linear activation function in both the encoder and decoder is ReLU. We use a linear layer with $[K,64]$ weights to map the latent variable $z^t$ to the same space of the ID features. \yhq{For $f_{cls}$ used in rating prediction, we first apply the ReLU function on the result of element-wise multiplication of the user and the item features, 
then apply a linear layer with the shape of $[64,1]$ to transform the ReLU output to a rating scalar.} 




 

\section{Transferable Information Bottleneck}
\label{app:theory}
\textbf{Theorem} Suppose we use independent well-trained encoder-decoder based architectures to model the features for different modalities (e.g., graph and text), then a given input $x_n$ has representations $x^t_n \in X^t$ and $x^g_n \in X^g$ under different modalities with the constraint $I(X^t;X^g) \geq I_x$ since $X^t$ and $X^g$ which are from the same input should be relevant. Also, assuming $I(Z^t;X^t) \leq I_c$ and $I(Z^g;X^g) \leq I_c$, then the following property holds:
\begin{equation}
    I(X^t;Z^g) \leq H(X^t) - H(X^g) + H(X^g|X^t) + I_c
\end{equation}

\noindent \textsl{Proof:} \lin{We first apply the chain rule twice in different orders for three sources $X^t$, $X^g$ and $Z^g$ below:}

\small{
\begin{gather}
   \lin{ H(X^t,X^g,Z^g) = H(X^g|X^t,Z^g) + H(X^t|Z^g) + H(Z^g),} \\ 
    \lin{H(X^t,X^g,Z^g) = H(X^t|X^g,Z^g) + H(X^g|Z^g) + H(Z^g). }
\end{gather}
}
\noindent \lin{Since the right parts of the above two equations are the same, we have:}
\begin{align}
    H(X^g|X^t,Z^g) + H(X^t|Z^g) + H(Z^g) \notag \\
    = H(X^t|X^g,Z^g) + H(X^g|Z^g) + H(Z^g)
\end{align}

\noindent \lin{which can be simplified as:}
\small{
\begin{equation}
   \lin{H(X^t|Z^g) = H(X^g|Z^g) + H(X^t|X^g,Z^g) - H(X^g|X^t,Z^g).}
\end{equation}
}

\noindent \lin{Since $Z^g$, $X^t$ and $X^g$ are derived from the same set of samples but under different modalities or representations, they are obviously not independent. Hence, we have $H(X^g|X^t,Z^g) \leq H(X^g|X^t)$ (as $X^g$ and $Z^g$ are dependent) and $H(X^t|X^g,Z^g) \geq 0$ (as $X^t$ and $X^g$ are dependent). Thus, we can replace  $H(X^g|X^t,Z^g)$ with $H(X^g|X^t)$, and replace $ H(X^t|X^g,Z^g)$ with 0, to derive the following inequality:}
\begin{equation}
    \lin{H(X^t|Z^g) \geq H(X^g|Z^g) - H(X^g|X^t)}
\end{equation}

\lin{By applying the properties of conditional differential entropy, which yields $H(X^g|Z^g) = H(X^g) - I(X^g;Z^g)$, the above formula can be simplified as: }

\begin{equation}
    \lin{H(X^t|Z^g) \geq H(X^g) - I(X^g;Z^g) - H(X^g|X^t)}
\end{equation}

Accordingly, we have:
\begin{align} 
    I(X^t;Z^g) &\lin{= H(X^t) - H(X^t|Z^g)}  \notag \\
    & \leq H(X^t) - H(X^g) + I(X^g;Z^g) + H(X^g|X^t)   \notag \\
    &\leq H(X^t) - H(X^g)  + H(X^g|X^t) + I_c \label{eq:MI}
\end{align} 
$\Box$

Eq~\ref{eq:MI} shows that the lowest upper bound of the transferable information bottleneck is $I(X^t;Z^g) \leq I_c$ when the distribution of two modalities are the same, $X^t = X^g$, and the conditional entropy of $H(X^g|X^t)$ is $0$. Let's take $H =  H(X^g) - H(X^t) + H(X^t|X^g)$, then the Eq. (\ref{eq:MI}) can be simplified as $I(X^g;Z^t) \leq H + I_c$, where $I_c$ is the upper boundary of information bottleneck for both modalities.

\lin{Recall the Lagrange multiplier based objective function: }

\begin{equation}
    \lin{O_{IB} = I(\Bar{X};Z) - \beta \cdot I(X;Z), }
\end{equation}

\lin{which aims to optimise:}

\begin{equation}
    \lin{\mathop{\rm{max}}\limits_{\theta} I(\Bar{X};Z) \quad {\rm s.t.} \quad I(X;Z)) \leq I_c.}
\end{equation}

\lin{Now, let us focus on the optimisation on the encoder-decoder framework in text by denoting the variables with the $t$ superscript. If we use the latent distribution learned from the graph modality, $Z^g$, to impose constraints on $Z^t$, the objective can be rewritten as:}

\begin{equation}
    \lin{\mathop{\rm{max}}\limits_{\theta} I(\Bar{X}^t;Z^t) \quad {\rm s.t.} \quad I(X^t;Z^g) \leq H + I_c.}
\end{equation}

\lin{Since $H =  H(X^g) - H(X^t) + H(X^t|X^g)$ is irrelevant with our optimisation, the Lagrange multiplier based objective function $O^t_{IB}$ can be rewritten as:}

\begin{equation}
    \lin{O^t_{IB} = I(\Bar{X}^t;Z^t) - \beta \cdot I(X^t;Z^g), }
\end{equation}

\lin{That is, there is no need to define the prior for the text modality. Instead, we can use the posterior distribution of the latent variable from the graph modality as the regularisation to guide the training on the text data. }

\section{Optimising Objective of Transferable Information Bottleneck}
\label{app:theory2}

\textbf{Information Bottleneck:} Recall that the IB objective in this work has the form $I(Z^t; \bar{X}^t) - \beta \cdot I(X^t;Z^g)$, where
 \begin{equation}
     I(Z^t; \bar{X}^t) = \int p(\bar{x}^t,z^t) \cdot {\rm log} \frac{p(\bar{x}^t|z^t)}{p(\Bar{x^t})} d\bar{x}^tdz^t, 
 \end{equation}
 where  $p(\cdot)$ is the true distribution which is not observed. According to the proof given by \cite{DBLP:conf/iclr/AlemiFD017}, we consider the posterior estimation $q(\bar{x}_t|z_t)$ in the decoder.  According to Gibbs' inequality, we have:
 \begin{align}
     I(Z^t; \bar{X}^t) &\geq \int p(\bar{x}^t,z^t) \cdot {\rm log} \frac{p(\bar{x}^t|z^t)}{q(\Bar{x^t})} d\bar{x}^tdz^t \\  \notag
     & = \int p(\bar{x}^t,z^t) \cdot {\rm log} {p(\bar{x}^t|z^t)} d\bar{x}^tdz^t - H(\Bar{X^t}). 
\end{align}

Since the entropy of the decoding results $H(\Bar{X^t})$ is independent of model optimisation, we only need to consider the posterior estimation based lower bound. According to\cite{DBLP:conf/iclr/AlemiFD017}, we have the following empirical approximation by reparameterisation trick:
\begin{equation}
     \mathbb{E}_{x^t \sim X^t} \mathbb{E}_{\epsilon \sim p(\epsilon)} [{\rm log}q(\Bar{x}^t|f(x^t,\epsilon)] 
\end{equation}

For the Lagrange multiplied term $I(X^t,Z^g)$ which is different from the standard information bottleneck based $\beta$-VAE, we do consider the information bottleneck of $I(X^t,Z^t)$ first, because $I(X^t,Z^g) \leq H + I(X^t,Z^t) $ according to the proof given in section \ref{app:theory}. Here, we have the following bound by applying Gibbs' inequality:

\begin{equation} \small
     I(X^t;Z^t) = \int p(x^t,z^t) {\rm log} p(z^t|x^t) dx^tdz^t - \int p(z^t) {\rm log} p(z^t) dz^t. 
\end{equation}
 
Since estimating the prior distribution of $Z_t$ might be difficult, based on the definition of transferable information bottleneck, we apply the Gibbs' inequality and have:

\begin{equation}
     I(X^t;Z^g) \leq \int p(z^t|x^t)p(x^t){\rm log}\frac{p(z^t|x^t)}{r(z^g)}dx^tdz^t + H,
\end{equation}

where $r(z^g)$ is the posterior distribution from the graph-based latent representation, and $H$ is decided by the prior of two modalities which is independent of model optimisation. 
By combining the two results, the above bound can be approximated by the reparameterisation trick \cite{https://doi.org/10.48550/arxiv.1312.6114} with a Gaussian random variable $\epsilon$:
\begin{align}
  O_{IB} &=  \mathbb{E}_{x^t \sim X^t}  \mathbb{E}_{\epsilon \sim p(\epsilon)}{[ - \underbrace{{\rm log}q(\Bar{x}^t|f(x^t,\epsilon)]}_{\rm Recons. \; term:\;{Q(x^t)}} } \notag \\
  &+ \beta \cdot\underbrace{ {\rm KL}(p(Z^t|x^t,\epsilon)|r(Z^g))}_{\rm KL-div. \; term: \; KL(Z^t|Z^g)},
\end{align}

\textbf{Further discussion about the reconstruction term $Q(x^t)$}

In section II,
we propose to use the Nadaraya-Watson estimator\cite{10.2307/2669691} to approximate the conditional probability $q(\Bar{x}^t|f(x^t,\epsilon)]$. The idea is to insert the decoding result $\Bar{x}^t$ to the space of training samples $\{x^t\}$, and take the kernel density estimation method to approximate $Q(x^t)$ by Bayesian rule: 

\begin{align}
    Q(x^t) = {\rm log}\frac{\hat{p}(\Bar{x}^t_{\epsilon},x^t)}{\hat{p}(x^t)} = {\rm log}\frac{\frac{1}{N} \sum_{j=1}^N \kappa(\frac{x^t-x^t_j}{h}) \cdot \kappa(\frac{\Bar{x}^t_{\epsilon}-x^t_j}{h})}{\frac{1}{N} \sum_{j=1}^N 
    \kappa(\frac{x^t-x^t_j}{h}) },
\end{align}

Since the input feature for ${x^t}$ is fixed, we only care about the updating of reconstruction through encoder-decoder architecture. Thus we have:

\begin{align}
   Q(x^t) &\varpropto  \frac{1}{N}\sum_{j=1}^N \kappa(\frac{x^t-x^t_j}{h}) \cdot \kappa(\frac{\Bar{x}^t_{\epsilon}-x^t_j}{h})
\end{align}

To simplify the above estimation, we apply the RBF-kernel and triangle inequality in estimation and have:

\begin{align}
\notag
   Q(x^t) &\varpropto  {\rm log}\frac{1}{N}\sum_{j=1}^N \kappa(\frac{x^t-x^t_j}{h}) \cdot  \kappa(\frac{\Bar{x}^t_{\epsilon}-x^t_j}{h}) \\ \notag
   &={\rm log}\frac{1}{N}\sum_{j=1}^N {\rm exp}(- ||x^t-x^t_j||^2) \cdot {\rm exp}(- ||\Bar{x}^t_{\epsilon}-x^t_j||^2) \\ \notag
   &={\rm log}\frac{1}{N}\sum_{j=1}^N {\rm exp}(- ||x^t-x^t_j||^2 - ||\Bar{x}^t_{\epsilon}-x^t_j||^2) \\ \notag
   &\leq {\rm log}\frac{1}{N}\sum_{j=1}^N {\rm exp}(- ||\Bar{x}^t_{\epsilon}-x^t||^2) \\ \notag
   &={\rm log}[{\rm exp}(- ||\Bar{x}^t_{\epsilon}-x^t||^2)] \\ 
\end{align}

Thus, we are able to optimise the $Q(x^t)$ by mean square error if we choose the natural logarithm function:

\begin{align}
    -Q(x^t) &\geq - {\rm ln} ({\rm exp}(- ||\Bar{x}^t_{\epsilon}-x^t||^2)) =||\Bar{x}^t_{\epsilon}-x^t||^2  
\end{align}

Therefore, we can rewrite the lower boundary as well as the learning objective $O_{IB}$ by:

\begin{align}
\small
  O_{IB} = \mathbb{E}_{x^t \sim X^t} \mathbb{E}_{\epsilon \sim p(\epsilon)} [||\Bar{x}^t_{\epsilon} - x^t||^2] + \beta \cdot {\rm KL}(p(Z^t|x^t,\epsilon)|r(Z^g)),
\end{align}

\section{Additional Experiment Results}
\yhq{We also compare with the generated clusters using \textsl{StandPrior}. The results are shown in Table~\ref{tab:mmd_wordcloud}. It is difficult to see a clear topic pattern in \textsl{StandPrior} as the top words largely overlap in different clusters. The first several words are all `\emph{hop}', `\emph{malt}' and `\emph{good}'. The results are even worse than \textsl{WassersteinVAE}, which can at least generate diverse clusters, reflected in the distinguished words in each cluster.}

\renewcommand{\arraystretch}{1.5}
\begin{table}[h]
\centering
\resizebox{0.48\textwidth}{!}{%
\begin{tabular}{p{1.cm}p{8.5cm}}
\toprule[2pt]
 \multirow{3}{*}{\textsl{Beer}} & hop, good, malt, aroma, \textcolor{black}{\textbf{hint}}, pour, \textcolor{black}{\textbf{great}}, well, mouthfeel, much\\
 \cline{2-2}
 & hop, aroma, good, sweet, light, malt, finish, \textcolor{black}{\textbf{brew}}, poured, \textcolor{black}{\textbf{glass}} \\
\cline{2-2}
  &hop, malt, light, sweet, much, \textcolor{black}{\textbf{carbonation}},  good, well, mouthfeel, pour \\
 \hline
\multirow{3}{*}{\textsl{Music}}
&great, best, band, love, \textcolor{black}{\textbf{hit}}, well, first, make, lyric, come\\
   \cline{2-2}
 &love, great, best, band, \textcolor{black}{\textbf{sound}}, well, first, even, make, lyric\\
 \cline{2-2}
&great, love, best,  well, band, lyric, make, fan, first \\
\hline
\multirow{3}{*}{\textsl{Office}}&
 use, work, well, binder, great, will, \textcolor{black}{\textbf{easy}}, \textcolor{black}{\textbf{nice}}, label, \textcolor{black}{\textbf{ink}} \\
 \cline{2-2}
&
label, easy, tape, color, great, using, work, make, well, really\\
 \cline{2-2}
&
 use, label, product, tape, color, binder, \textcolor{black}{\textbf{quality}}, really, \textcolor{black}{\textbf{scanner}}, great\\
\bottomrule[2pt]
    \end{tabular}}
    \caption{The most prominent words (sorted by occurrence frequency) in three randomly selected clusters from \textsl{StandPrior} on \textsl{BeerAdvocate}, \textsl{Digital Music} and \textsl{Office Products} datasets. For each dataset, we highlight the top 3 words that are not found in the other two clusters. \textsl{StandPrior} produces topics which contain largely overlapped words.}
    \label{tab:mmd_wordcloud}
\vspace{-10pt}
\end{table}
\end{document}